\documentclass[letterpaper,twocolumn,10pt]{article}
\usepackage{usenix,epsfig,endnotes,balance,epsfig,graphicx,amsmath,amssymb}
\usepackage{amsfonts,color,multirow,algorithm,algorithmic,epstopdf,hhline}
\usepackage{tabularx,subcaption} 
\usepackage[hyphens]{url}
\usepackage[T1]{fontenc}
\usepackage[table]{xcolor}

\newcommand{\para}[1]{{\vspace{4pt} \bf \noindent #1 \hspace{10pt}}}
\newcommand{\parab}[1]{{\vspace{4pt} \bf \noindent #1}}

\newcommand{\diff}[1]{{#1}}

\newenvironment{packed_itemize}{
\begin{itemize}
 \setlength{\itemsep}{2pt}
 \setlength{\parskip}{0pt}
 \setlength{\parsep}{0pt}
 \setlength{\headsep}{0pt}
 \setlength{\topskip}{0pt}
 \setlength{\topmargin}{0pt}
 \setlength{\topsep}{0pt}
 \setlength{\partopsep}{0pt}
}{\end{itemize}}

\newenvironment{packed_enumerate}{
\begin{enumerate}
  \setlength{\itemsep}{1pt}
  \setlength{\parskip}{0pt}
  \setlength{\parsep}{0pt}
  \setlength{\headsep}{0pt}
  \setlength{\topskip}{0pt}
  \setlength{\topmargin}{0pt}
  \setlength{\topsep}{0pt}
  \setlength{\partopsep}{0pt}
}{\end{enumerate}}

\newcommand{\eg}{{\it e.g.,\ }}
\newcommand{\etal}{{\it et al.\ }}

\newcommand{\ie}{{\it i.e.,\ }}

\begin{document}

\date{}

\title{\Large \bf Tracing Information Flows Between Ad Exchanges Using Retargeted Ads}

\author{
{\rm Muhammad Ahmad Bashir}\\
Northeastern University\\
ahmad@ccs.neu.edu
\and
{\rm Sajjad Arshad}\\
Northeastern University\\
arshad@ccs.neu.edu
\and
{\rm William Robertson}\\
Northeastern University\\
wkr@ccs.neu.edu
\and
{\rm Christo Wilson}\\
Northeastern University\\
cbw@ccs.neu.edu
}

\maketitle
\sloppy

\subsection*{Abstract}

Numerous surveys have shown that Web users are concerned about
the loss of privacy associated with online tracking. Alarmingly, these surveys also reveal
that people are also unaware of the amount of data sharing that occurs between ad exchanges,
and thus underestimate the privacy risks associated with online tracking.

In reality, the modern ad ecosystem is fueled by a flow of user data between trackers
and ad exchanges. Although recent work has shown that ad exchanges routinely perform
{\em cookie matching} with other exchanges, these studies are based on brittle heuristics
that cannot detect all forms of information sharing, especially under adversarial conditions.

In this study, we develop a methodology that is able to detect client- and server-side flows of
information between arbitrary ad exchanges. Our key insight is to leverage {\em retargeted ads}
as a tool for identifying information flows. Intuitively, our methodology works because
it relies on the {\em semantics} of how exchanges serve ads, rather than focusing on specific
cookie matching {\em mechanisms}. Using crawled data on 35,448 ad impressions, we show that our
methodology can successfully categorize four different kinds of information sharing behavior between
ad exchanges, including cases where existing heuristic methods fail.

\diff{We conclude with a discussion of how our findings and methodologies can be leveraged to 
give users more control over what kind of ads they see and how their information is shared between ad exchanges.}

\section{Introduction}
\label{sec:intro}

\diff{
People have complicated feelings with respect to online behavioral advertising. While surveys
have shown that some users prefer relevant, targeted ads to random, untargeted
ads~\cite{ur-soups12,chanchary-soups15}, this preference has caveats. For example, 
users are uncomfortable with ads that are targeted based on sensitive Personally Identifiable
Information (PII)~\cite{malheiros-chi12,agarwal-soups13} or specific kinds of browsing history
(\eg visiting medical websites)~\cite{leon-soups13}. Furthermore, some users are universally
opposed to online tracking, regardless of
circumstance~\cite{mcdonald-wpes10,ur-soups12,chanchary-soups15}.

One particular concern held by users is their ``digital footprint''~\cite{howell-tr15,wolpin-hp15,spector-pcw16},
\ie which first- and
third-parties are able to track their browsing history? Large-scale web crawls have repeatedly shown
that trackers are ubiquitous~\cite{gill-imc13,englehardt-www15}, with DoubleClick alone being able
to observe visitors on 40\% of websites in the Alexa Top-100K~\cite{cahn-www16}. These results
paint a picture of a balkanized web, where trackers divide up the space and compete
for the ability to collect data and serve targeted ads.

However, this picture of the privacy landscape is at odds with the current reality
of the ad ecosystem. Specifically, ad exchanges routinely perform
{\em cookie matching} with each other, to synchronize unique identifiers and share
user data~\cite{acar-WNF-2014,olejnik-ndss14,falahrastegar-pam16}. Cookie matching
is a precondition for ad exchanges to participate in {\em Real Time Bidding} (RTB) auctions,
which have become the dominant mechanism for buying and selling advertising inventory from
publishers. Problematically, Hoofnagle \etal report that users na\"{\i}vely believe
that privacy policies prevent companies from sharing user
data with third-parties, which is not always the case~\cite{hoofnagle-wflr14}.
}

Despite user concerns about their digital footprint, we currently lack the tools to
fully understand how information is being shared between ad exchanges. Prior empirical work on
cookie matching has relied on heuristics that look for specific strings in HTTP messages to identify
flows between ad networks~\cite{acar-WNF-2014,olejnik-ndss14,falahrastegar-pam16}. However, these
heuristics are brittle in the face of obfuscation: for example, DoubleClick cryptographically
hashes their cookies before sending them to other ad networks~\cite{doubleclick-rtb}. More
fundamentally, analysis of {\em client-side} HTTP messages are insufficient to detect
{\em server-side} information flows between ad networks.

\break

In this study, we develop a methodology that is able to detect client- and server-side flows of
information between arbitrary ad exchanges \diff{that serve {\em retargeted ads}. Retargeted ads
are the most specific form of behavioral ads, where a user is targeted with ads related to the exact
products she has previously browsed (see \S~\ref{sec:targets} for definition). For example, Bob
visits \url{nike.com} and browses for running shoes but decides not to purchase them. Bob later 
visits \url{cnn.com} and sees an ad for the exact same running shoes from Nike.} 

Our key insight is to leverage retargeted ads as a mechanism for identifying
information flows. This is possible because the strict conditions that must be met for a
retarget to be served 
allow us to infer the precise flow of tracking information that facilitated the serving of the
ad. Intuitively, our methodology works because it relies on the {\em semantics} of how
exchanges serve ads, rather than focusing on specific cookie matching {\em mechanisms}. 

To demonstrate the efficacy of our methodology, we conduct extensive experiments on real data.
We train 90 {\em personas} by visiting popular e-commerce sites, and then crawl major publishers
to gather retargeted ads~\cite{barford-adscape-2014,carrascosa-conext15}. Our crawler is an
instrumented version of Chromium that records the {\em inclusion chain} for every resource it
encounters~\cite{arshad-fc16}, including 35,448 chains associated with 5,102 unique retargeted ads. 
We use carefully designed pattern matching rules to categorize each of these chains, which reveal
1) the pair of ad exchanges that shared information in order to serve the retarget, and 2) the
mechanism used to share the data (\eg cookie matching).

In summary, we make the following contributions:
\begin{packed_itemize}
\item We present a novel methodology for identifying information flows between ad networks that is
    content- and ad exchange-agnostic. Our methodology allows to identify four different categories
    of information sharing between ad exchanges, of which cookie matching is one.
\item Using crawled data, we show that the heuristic methods used by prior work to analyze cookie matching
    are unable to identify 31\% of ad exchange pairs that share data.
\item Although it is known that Google's privacy policy allows it to share data between its
    services~\cite{google-tos}, we provide the first empirical evidence that Google uses this capability to
    serve retargeted ads.
\item Using graph analysis, we show how our data can be used to automatically infer the roles played
    by different ad exchanges (\eg Supply-Side and Demand-Side Platforms). These results expand upon
    prior work~\cite{gomer-wiiat13} and facilitate a more nuanced understanding of the online ad ecosystem.
\end{packed_itemize}

Ultimately, we view our methodology as a stepping stone towards more balanced privacy protection
tools for users, that also enable publishers to earn revenue. \diff{Surveys have shown that users are
not necessarily opposed to online ads: some users are just opposed to tracking~\cite{mcdonald-wpes10,ur-soups12,chanchary-soups15},
while others simply desire more nuanced control over their digital footprint~\cite{agarwal-soups13,leon-soups13}.
However, existing tools (\eg browser extensions) cannot
distinguish between targeted and untargeted ads, thus leaving
users with no alternative but to block all ads.} Conversely, our results open up the possibility
of building in-browser tools that just block cookie matching, which will effectively prevent most
targeted ads from RTB auctions, while still allowing untargeted ads to be served. 

\para{Open Source.} As a service to the community, we have open sourced all the data
from this project. This includes over 7K labeled behaviorally targeted and
retargeted ads, as well as the inclusion chains and full HTTP traces associated with these ads.
The data is available at:
\begin{center}
\vspace{-0.5mm}
{\tt http://personalization.ccs.neu.edu/}
\vspace{-3mm}
\end{center}

\begin{figure*}[t]
	\centering
	\includegraphics[width=1\textwidth]{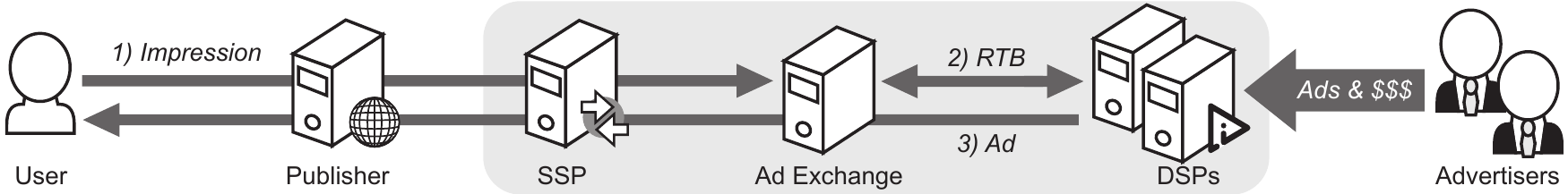}
	\caption{The display advertising ecosystem. Impressions and tracking data flow left-to-right, while revenue and ads flow right-to-left.}
	\label{fig:adnetworks}
\end{figure*}

\begin{figure}[b]
	\centering
	\includegraphics[width=1\columnwidth]{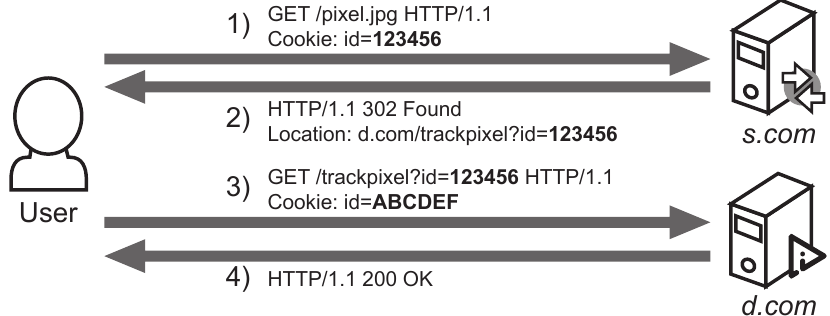}
	\caption{SSP $s$ matches their cookie to DSP $d$ using an HTTP redirect.}
	\label{fig:cookiematch}
\end{figure}

\section{Background and Definitions}
\label{sec:bg}

In this section, we set the stage for our study by providing background about the online display
ad industry, as well as defining key terminology. We focus on techniques and terms related to 
Real Time Bidding and retargeted ads, since they are the focus of our study.

\subsection{Online Display Advertising}

Online display advertising is fundamentally a matching problem. On one side are
{\em publishers} (\eg news websites, blogs, {\em etc.}) who produce content, and earn revenue
by displaying ads to users. On the other side are advertisers who want to display ads to
particular users (\eg based on demographics or market segments). Unfortunately, the online
user population is fragmented across hundreds of thousands of publishers, making it difficult
for advertisers to reach desired customers.

{\em Ad networks} bridge this gap by aggregating {\em inventory} from publishers (\ie space
for displaying ads) and filling it with ads from advertisers. Ad networks make it possible for
advertisers to reach a broad swath of users, while also guaranteeing a steady stream of revenue
for publishers. Inventory is typically sold using a Cost per Mille (CPM) model, where
advertisers purchase blocks of 1000 {\em impressions} (views of ads), or a Cost per Click (CPC)
model, where the advertiser pays a small fee each time their ad is clicked by a user. 

\para{Ad Exchanges and Auctions.} Over time, ad networks are being supplanted by
{\em ad exchanges} that rely on an auction-based model. In Real-time Bidding ({\em RTB}) exchanges,
advertisers bid on individual impressions, in real-time; the winner of the auction is permitted
to serve an ad to the user. Google's DoubleClick is the largest ad exchange, and it supports RTB.

As shown in Figure~\ref{fig:adnetworks}, there is a distinction between Supply-side Platforms
({\em SSPs}) and Demand-side Platforms ({\em DSPs}) with respect to ad auctions. SSPs work
with publishers to manage their relationships with multiple ad exchanges, typically to maximize
revenue. For example, OpenX is an SSP. In contrast, DSPs work with advertisers to assess the
value of each impression and optimize bid prices. MediaMath is an example of a DSP.
To make matters more complicated, many companies offer products that cross categories;
for example, Rubicon Project offers SSP, ad exchange, and DSP products. We direct interested readers to
\cite{mayer-2012-TWT} for more discussion of the modern online advertising ecosystem.

\subsection{Targeted Advertising}
\label{sec:targets}

Initially, the online display ad industry focused on generic brand ads (\eg ``Enjoy Coca-Cola!'')
or {\em contextual ads} (\eg an ad for Microsoft on StackOverflow). However, the industry quickly
evolved towards {\em behavioral targeted ads} that are served to specific users
based on their browsing history, interests, and demographics.

\para{Tracking.} To serve targeted ads, ad exchanges and advertisers must collect data about online
users by tracking their actions. Publishers embed JavaScript or invisible ``tracking pixels'' that
are hosted by tracking companies into their web pages, thus any user who visits the publisher also
receives third-party cookies from the tracker (we discuss other tracking mechanisms in
\S~\ref{sec:related}). Numerous studies have shown that trackers are pervasive across the
Web~\cite{krishnamurthy-imc06,krishnamurthy-www09,roesner-2012-DDA,cahn-www16},
which allows advertisers to collect users' browsing history.
All major ad exchanges, like DoubleClick and Rubicon, perform user tracking,
but there are also companies like BlueKai that just specialize in tracking. 

\para{Cookie Matching.} During an RTB ad auction, DSPs submit bids on an impression. The amount
that a DSP bids on a given impression is intrinsically linked to the amount of information they have
about that user. For example, a DSP is unlikely to bid highly for user $u$ whom they have never observed
before, whereas a DSP may bid heavily for user $v$ who they have recently observed browsing high-value
websites (\eg the baby site \url{TheBump.com}).

However, the Same Origin Policy (SOP) hinders the ability of DSPs to identify users in ad auctions. As shown
in Figure~\ref{fig:adnetworks}, requests are first sent to an SSP which forwards the impression to an exchange
(or holds the auctions itself). At this point,
the SSP's cookies are known, but not the DSPs. This leads to a catch-22 situation: a DSP cannot read its
cookies until it contacts the user, but it cannot contact the user without first bidding and winning the
auction.

To circumvent SOP restrictions, ad exchanges and advertisers engage in {\em cookie matching} (sometimes
called {\em cookie syncing}). Cookie matching is illustrated in Figure~\ref{fig:cookiematch}: the user's
browser first contacts ad exchange \texttt{s.com}, which returns an HTTP redirect to its partner
\texttt{d.com}. \texttt{s} reads its own cookie, and includes it as a parameter
in the redirect to \texttt{d}. \texttt{d} now has a mapping from its cookie to \texttt{s}'s. In
the future, if \texttt{d} participates in an auction held by \texttt{s}, it will be able to identify
matched users using \texttt{s}'s cookie.
Note that some ad exchanges (including DoubleClick) send cryptographically
hashed cookies to their partners, which prevents the ad network's true cookies from leaking to third-parties.

\para{Retargeted Ads.} In this study, we focus on {\em retargeted ads}, which are the most
specific type of targeted display ads. Two conditions must be met for a DSP to
serve a retargeted ad to a user $u$: 1) the DSP must know that $u$ browsed a specific
product on a specific e-commerce site, and 2) the DSP must be able to uniquely identify $u$ during
an auction. If these conditions are met, the DSP can serve $u$ a
highly personalized ad reminding them to purchase the product from the retailer. Cookie matching
is crucial for ad retargeting, since it enables DSPs to meet requirement (2).

\section{Related Work}
\label{sec:related}

Next, we briefly survey related work on online advertising. We begin by looking at more general studies
of the advertising and tracking ecosystem, and conclude with a more focused examination of studies on
cookie matching and retargeting. Although existing studies on cookie matching demonstrate that this
practice is widespread and that the privacy implications are alarming, these works have significant
methodological shortcomings that motivate us to develop new techniques in this work.

\subsection{Measuring the Ad Ecosystem}

Numerous studies have measured and broadly characterized the
online advertising ecosystem. Guha \etal were the first to systematically measure online ads, and
their carefully controlled methodology has been very influential on subsequent studies (including this
one)~\cite{guha-IMC-2010}. Barford \etal take a much broader look at the {\em adscape} to determine who the major
ad networks are, what fraction of ads are targeted, and what user characteristics drive
targeting~\cite{barford-adscape-2014}. Carrascosa \etal take an even finer grained look at targeted ads
by training {\em personas} that embody specific interest profiles (\eg cooking, sports), and find that
advertisers routinely target users based on sensitive attributes (\eg religion)~\cite{carrascosa-conext15}.
Rodriguez \etal measure the ad ecosystem on mobile devices~\cite{rodriguez-imc12}, while Zarras \etal
analyzed malicious ad campaigns and the ad networks that serve them~\cite{zarras-imc14}.

Note that {\bf none} of these studies examine retargeted ads; Carrascosa \etal specifically excluded retargets
from their analysis~\cite{carrascosa-conext15}.

\para{Trackers and Tracking Mechanisms.} To facilitate ad targeting, participants in the ad ecosystem
must extensively track users. Krishnamurthy \etal have been cataloging the spread of trackers and assessing
the ensuing privacy implications for years~\cite{krishnamurthy-imc06,krishnamurthy-www09,krishnamurthy-w2sp11}.
Roesner \etal develop a comprehensive taxonomy of different tracking mechanisms
that store state in users' browsers (\eg cookies, HTML5 LocalStorage, and Flash LSOs),
as well as strategies to block them~\cite{roesner-2012-DDA}. Gill \etal use large web browsing traces
to model the revenue earned by different trackers (or {\em aggregators} in their terminology), and found 
that revenues are skewed towards the largest trackers (primarily Google)~\cite{gill-imc13}.
More recently, Cahn \etal performed
a broad survey of cookie characteristics across the Web, and found that less than 1\% of trackers can aggregate
information across 75\% of websites in the Alexa Top-10K~\cite{cahn-www16}. Falahrastegar \etal expand on these
results by comparing trackers across geographic regions~\cite{falahrastegar-tma2014}, while Li \etal show that
most tracking cookies can be automatically detected using simple machine learning methods~\cite{li-pam15}.

Note that {\bf none} of these studies examine cookie matching, or information sharing between ad exchanges.

Although users can try to evade trackers by clearing their cookies or using private/incognito browsing modes,
companies have fought back using techniques like {\em Evercookies} and {\em fingerprinting}. Evercookies
store the tracker's state in many places within the browser (\eg FlashLSOs, etags, {\em etc.}), thus
facilitating regeneration of tracking identifiers even if users delete their
cookies~\cite{samykamkar-evercookies,soltani2010flash,ayenson2011flash,mcdonald-isjlp11}. Fingerprinting
involves generating a unique ID for a user based on the characteristics of their
browser~\cite{eckersley-2010-UYW,MBYS11,mulazzani-w2sp13}, browsing history~\cite{olejnik-hotpets12},
and computer (\eg the
HTML5 canvas~\cite{mowery2012pixel}). Several studies have found trackers in-the-wild that use
fingerprinting techniques~\cite{acar-2013-FDW,nikiforakis-2013-CME,kohno-2015-fingerprinting};
Nikiforakis \etal propose to stop fingerprinting by carefully and intentionally adding more entropy
to users' browsers~\cite{nikiforakis-www15}.

\para{User Profiles.} Several studies specifically focus on tracking data collected by Google, since
their trackers are more pervasive than any others on the Web~\cite{gill-imc13,cahn-www16}.
Alarmingly, two studies have found that Google's Ad Preferences Manager,
which is supposed to allow users to see and adjust how they are being targeted for ads,
actually hides sensitive information from users~\cite{wills-2012-wpes,datta-2015-adfisher}. This finding
is troubling given that several studies rely on data from the Ad Preferences Manager
as their source of ground-truth~\cite{guha-IMC-2010,castelluccia-pets12,barford-adscape-2014}. To
combat this lack of transparency, Lecuyer \etal have built systems that rely on controlled experiments
and statistical analysis to infer the profiles that Google constructs about
users~\cite{lecuyer-2014-xray,lecuyer-2015-sunlight}. Castelluccia \etal go further by showing
that adversaries can infer users' profiles by passively observing the targeted ads they are shown by
Google~\cite{castelluccia-pets12}.

\subsection{Cookie Matching and Retargeting}
\label{sec:backcookies}

Although ad exchanges have been transitioning to RTB auctions since the mid-2000s, only three empirical
studies have examined the cookie matching that enables these services. Acar \etal found that hundreds of
domains passed unique identifiers to each other while crawling websites in the Alexa Top-3K~\cite{acar-WNF-2014}. 
Olejnik \etal noticed that ad auctions were leaking the winning bid prices for impressions,
thus enabling a fascinating behind-the-scenes look at RTB auctions~\cite{olejnik-ndss14}. In addition
to examining the monetary aspects of auctions, Olejnik \etal found 125 ad exchanges using cookie matching.
Finally, Falahrastegar \etal examine the clusters of domains that all share unique, matched cookies
using crowdsourced browsing data~\cite{falahrastegar-pam16}. Additionally, Ghosh \etal use game theory
to model the incentives for ad exchanges to match cookies with their competitors, but they provide no
empirical measurements of cookie matching~\cite{ghosh-2012-MME}.

Several studies examine retargeted ads, which are directly facilitated by cookie matching
and RTB. Liu \etal identify and measure retargeted ads served by DoubleClick by relying on unique AdSense
tags that are embedded in ad URLs~\cite{liu-2013-AIT}. Olejnik \etal crawled specific e-commerce sites
in order to elicit retargeted ads from those retailers, and observe that retargeted ads can cost
advertisers over \$1 per impression (an enormous sum, considering contextual ads sell for
$<$\$0.01)~\cite{olejnik-ndss14}. 

\para{Limitations.} The prior work on cookie matching demonstrates that this practice is
widespread. However, these studies also have significant methodological limitations, which prevent
them from observing all forms of information sharing between ad exchanges. Specifically:
\begin{packed_enumerate}
\item All three studies identify cookie matching by locating unique user IDs that are transmitted
    to multiple third-party domains~\cite{acar-WNF-2014,olejnik-ndss14,falahrastegar-pam16}.
    Unfortunately, this will miss cases where exchanges send permuted or obfuscated IDs to their
    partners. Indeed, DoubleClick is known to do this~\cite{doubleclick-rtb}.
\item The two studies that have examined the behavior of DoubleClick have done so by relying on specific
    cookie keys and URL parameters to detect cookie matching and retargeting~\cite{olejnik-ndss14,liu-2013-AIT}.
    Again, these methods are not robust to obfuscation or encryption that hide the content of HTTP messages.
\item Existing studies cannot determine the precise information flows between ad exchanges, \ie which parties
    are sending or receiving information~\cite{acar-WNF-2014}. This limitation stems from analysis
    techniques that rely entirely
    on analyzing HTTP headers. For example, a script from \texttt{t1.com} embedded in \texttt{pub.com} may
    share cookies with \texttt{t2.com} using dynamic AJAX, but the referrer appears to be \url{pub.com}, thus
    potentially hiding \texttt{t1}'s role as the source of the flow. 
\end{packed_enumerate}
In general, these limitations stem from a reliance on analyzing specific {\em mechanisms} for cookie matching.
In this study, one of our primary goals is to develop a methodology for detecting cookie matching that
is agnostic to the underlying matching mechanism, and instead relies on the fundamental {\em semantics} of
ad exchanges.

\section{Methodology}
\label{sec:meth}

In this study, our primary goal is to develop a methodology for detecting flows of user data between
arbitrary ad exchanges. This includes client-side flows (\ie cookie matching), as
well as server-side flows.

In this section, we discuss the methods and data we use to meet this goal. First, we briefly sketch
our high-level approach, and discuss key enabling insights. Second, we introduce the instrumented version
of Chromium that we use during our crawls. Third, we explain how we designed and
trained shopper {\em personas} that view products on the web, and finally we detail how we collected ads using
the trained personas.

\subsection{Insights and Approach}

Although prior work has examined information flow between ad exchanges, these studies are limited to
specific types of cookie matching that follow well-defined patterns (see \S~\ref{sec:backcookies}).
To study arbitrary information flows in a mechanism-agnostic way, we need a fundamentally different
methodology.

We solve this problem by relying on a key insight: in most cases, if a user is served a retargeted
ad, this proves that ad exchanges shared information about the user \diff{(see \S~\ref{sec:cat_rules})}. 
To understand this insight, consider that two preconditions must be met for user $u$ to be served a retarget 
ad for $shop$ by DSP $d$. {\em First}, either $d$ directly observed $u$ visiting $shop$, or $d$ must be told 
this information by SSP $s$. If this condition is not met, then $d$ would not pay the premium price necessary 
to serve $u$ a retarget. {\em Second}, if the retarget was served from an ad auction, SSP $s$ and $d$ must be
sharing information about $u$. If this condition is not met, then $d$ would have no way of identifying
$u$ as the source of the impression (see \S~\ref{sec:targets}).

In this study, we leverage this observation to reliably infer information flows between SSPs and DSPs,
regardless of whether the flow occurs client- or server-side. The high-level methodology is quite intuitive:
have a clean browser visit specific e-commerce sites, then crawl publishers and gather ads. If we observe
retargeted ads, we know that ad exchanges tracking the user on the {\em shopper-side} are sharing information
with exchanges serving ads on the {\em publisher-side}. Specifically, our methodology uses the following steps:
\begin{packed_itemize}
\item{\S~\ref{subsec:chrome}:} We use an instrumented version of Chromium to record {\em inclusion chains}
    for all resources encountered during our crawls~\cite{arshad-fc16}. These chains record the precise origins
    of all resource requests, even when the requests are generated dynamically by JavaScript or Flash. We use
    these chains in \S~\ref{sec:analysis} to categorize information flows between ad exchanges.
\item{\S~\ref{sec:personas}:} To elicit retargeted ads from ad exchanges, we design {\em personas}
    (to borrow terminology from~\cite{barford-adscape-2014} and~\cite{carrascosa-conext15}) that visit
    specific e-commerce sites. These sites
    are carefully chosen to cover different types of products, and include a wide variety of common trackers. 
\item{\S~\ref{subsec:ad_collection}:} To collect ads, our personas crawl 150 publishers from the Alexa Top-1K.
\item{\S~\ref{sec:labeling}:} We leverage well-known filtering techniques and crowdsourcing to
    identify retargeted ads from our corpus of 571,636 unique crawled images.
\end{packed_itemize}

\begin{figure}[t]
   \centering
   \includegraphics[width=1.0\columnwidth]
   {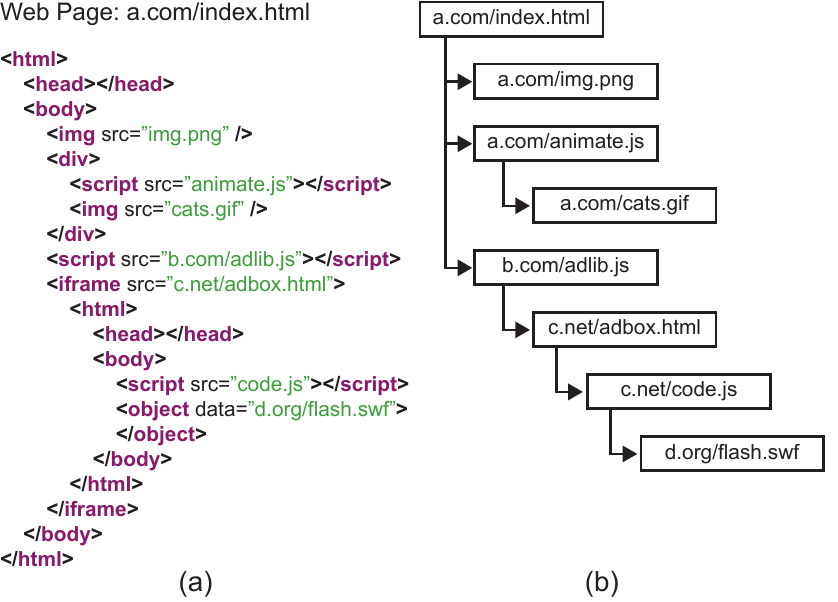}
   \caption{(a) DOM Tree, and (b) Inclusion Tree.}
   \label{fig:dom-inclusion-tree}
\end{figure}

\subsection{Instrumenting Chromium}
\label{subsec:chrome}

Before we can begin crawling, we first need a browser that is capable of recording detailed information about
the provenance of third-party resource inclusions in webpages. Recall that prior work on cookie matching
was unable to determine which ad exchanges were syncing cookies in many cases because the analysis relied
solely on the contents of HTTP requests~\cite{acar-WNF-2014,falahrastegar-pam16} (see \S~\ref{sec:backcookies}).
The fundamental problem is that HTTP requests, and even the DOM tree itself, do not reveal the true sources
of resource inclusions in the presence of dynamic code (JavaScript, Flash, {\em etc.}) from third-parties.

To understand this problem, consider the example DOM tree for \texttt{a.com/index.html} in
Figure~\ref{fig:dom-inclusion-tree}(a). Based on the DOM, we might conclude that the chain
$a \rightarrow c \rightarrow d$ captures the sequence of inclusions leading from the root of the page
to the Flash object from \texttt{d.org}.

However, direct use of a webpage's DOM is misleading because the DOM does not reliably
record the inclusion relationships between resources in a page. This is due to the ability of JavaScript to
manipulate the DOM at run-time, \ie by adding new inclusions dynamically. As such, while the DOM is a faithful
syntactic description of a webpage \emph{at a given point in time}, it cannot be relied upon to extract 
relationships between included resources. Furthermore, analysis of HTTP request headers does not solve this
problem, since the \texttt{Referer} is set to the first-party domain even when inclusions are dynamically added by
third-party scripts.

To solve this issue, we make use of a heavily instrumented version of Chromium that produces {\em inclusion trees}
directly from Chromium's resource loading code~\cite{arshad-fc16}. Inclusion trees capture the semantic
inclusion structure of resources in a webpage (\ie which objects cause other objects to be loaded), unlike
DOM trees which only capture syntactic structures. \diff{Our instrumented Chromium accurately captures
relationships between elements, regardless of where they are located (\eg within a single page or across frames) 
or how the relevant code executes (\eg via an inline \texttt{<script>}, \texttt{eval()}, or
an event handler).} We direct interested readers to \cite{arshad-fc16}
for more detailed information about inclusion trees, and the technical details of how the Chromium binary
is instrumented.

Figure~\ref{fig:dom-inclusion-tree}(b) shows the inclusion tree corresponding to the DOM tree in
Figure~\ref{fig:dom-inclusion-tree}(a). From the inclusion
tree, we can see that the true {\em inclusion chain} leading to the Flash object is
$a \rightarrow b \rightarrow c \rightarrow c \rightarrow d$, since the \texttt{IFrame} and the Flash are
dynamically included by JavaScript from \texttt{b.com} and \texttt{c.net}, respectively.

Using inclusion chains, we can precisely analyze the provenance of third-party resources included in webpages.
In \S~\ref{sec:analysis}, we use this capability to distinguish client-side flows of information between ad
exchanges (\ie cookie matching) from server-side flows.

\begin{figure}[t]
\begin{center}
\includegraphics[width=1.0\columnwidth]
{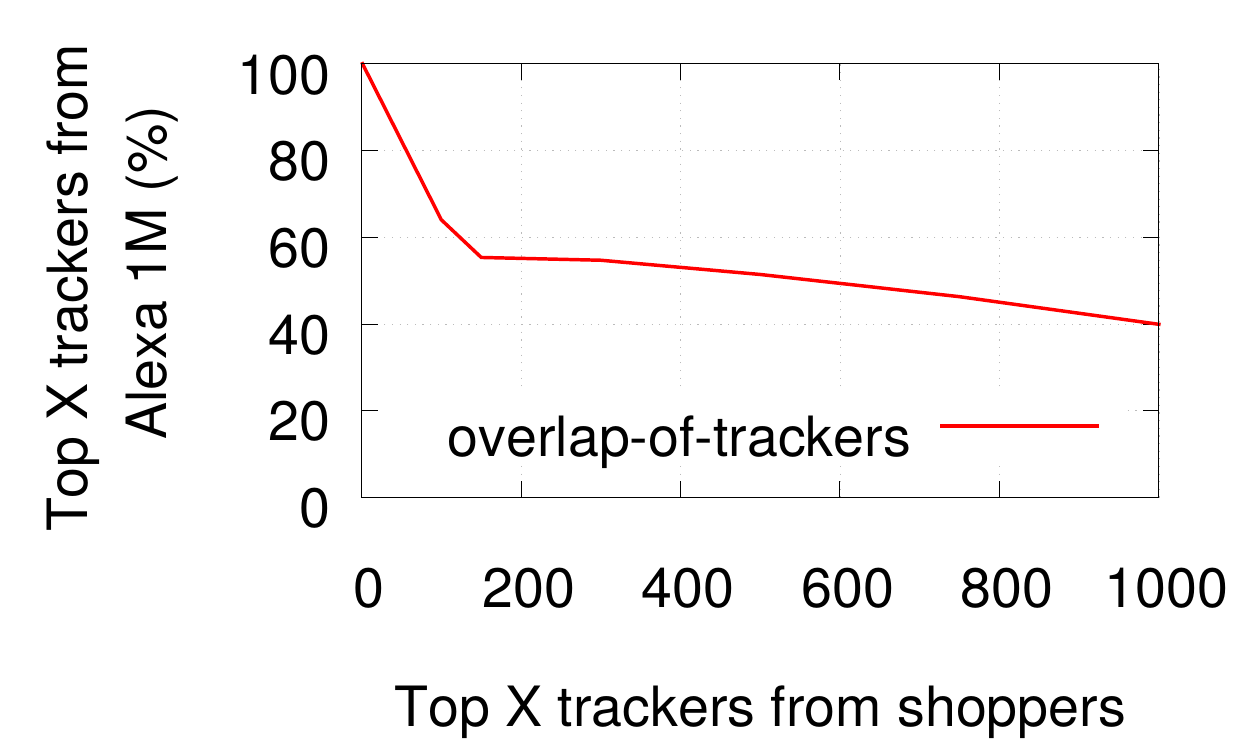}
\end{center}
\caption{Overlap between frequent trackers on e-commerce sites and Alexa Top-5K sites.}
\label{fig:tracker_overlap}
\end{figure}

\subsection{Creating Shopper Personas}
\label{sec:personas}

Now that we have a robust crawling tool, the next step in our methodology is designing shopper personas.
Each persona visits products on specific e-commerce sites, in hope of seeing retargeted
ads when we crawl publishers.

Since we do not know a priori which e-commerce sites are conducting retargeted ad campaigns, our personas
must cover a wide variety of sites. To facilitate this, we leverage the hierarchical categorization
of e-commerce sites maintained by Alexa\footnote{\url{http://www.alexa.com/topsites/category/Top/Shopping}}.
Although Alexa's hierarchy has 847 total categories, there is significant overlap between categories. We
manually selected 90 categories to use for our personas that have minimal overlap, as well as cover major
e-commerce sites (\eg Amazon and Walmart) and shopping categories (\eg sports and jewelry).

For each persona, we included the top 10 e-commerce sites in the corresponding Alexa category. In total, the
personas cover 738 unique websites. Furthermore, we manually selected 10 product URLs on each
of these websites. Thus, each persona visits 100 products URLs.

\para{Sanity Checking.} The final step in designing our personas is ensuring that the e-commerce sites
are embedded with a representative set of trackers. If they are not, we will not be able to collect targeted ads.

Figure~\ref{fig:tracker_overlap} plots the overlap between the trackers we observe on the Alexa Top-5K websites,
compared to the top $x$ trackers (\ie most frequent) we observe on the e-commerce sites. We see that 84\% of
the top 100 e-commerce trackers
are also present in the trackers on Alexa Top-5K sites\footnote{\diff{We separately crawled the resources 
included by the Alexa Top-5K websites in January 2015. For each website, we visited 6 pages and recorded all the 
requested resources.}}. 
These results demonstrate that our shopping personas will be seen by the vast majority of major 
trackers when they visit our 738 e-commerce sites.

\subsection{Collecting Ads}
\label{subsec:ad_collection}

In addition to selecting e-commerce sites for our personas, we must also select publishers to crawl for ads.
We manually select 150 publishers by examining the Alexa Top-1K websites and filtering out those which do
not display ads, are non-English, are pornographic, or require logging-in to view content (\eg Facebook).
We randomly selected 15 URLs on each publisher to crawl.

At this point, we are ready to crawl ads. We initialized 91 copies of our instrumented Chromium binary: 90
corresponding to our shopper personas, and one which serves as a control. During each {\em round} of
crawling, the personas visit their associated e-commerce sites, then visit the 2,250 publisher URLs
(150 publishers $*$ 15 pages per publisher). The control {\em only} visits the publisher URLs, \ie
it does not browse e-commerce sites, and therefore should never be served retargeted ads. The crawlers 
are executed in tandem, so they visit the publishers URLs in the same order at the same times. We hard-coded
a 1 minute delay between subsequent page loads to avoid overloading any servers, and to allow time for
the crawler to automatically scroll to the bottom of each page. Each round takes 40 hours to complete.

We conducted nine rounds of crawling between December 4 to 19, 2015. We stopped after 9 rounds because we
observed that we only gathered 4\% new images during the ninth round. The crawlers recorded inclusion trees,
HTTP request and response headers, cookies, and images from all pages. At no point did our crawlers click on
ads, \diff{since this can be construed as click-fraud (\ie advertisers often have to pay each time their ads are
clicked, and thus automated clicks drain their advertising budget). All crawls were done from {\em Northeastern 
University's} IP addresses in Boston.}

\section{Image Labeling}
\label{sec:labeling}

Using the methodology in \S~\ref{subsec:ad_collection}, we collected 571,636 unique images in total. However,
only a small subset are retargeted ads, which are our focus.
In this section, we discuss the steps we used to filter down our image set and isolate retargeted ads,
beginning with standard filters used by prior work~\cite{barford-adscape-2014,li-pam15},
and ending with crowdsourced image labeling.

\begin{figure}[t]
\begin{center}
\includegraphics[width=1.0\columnwidth]
{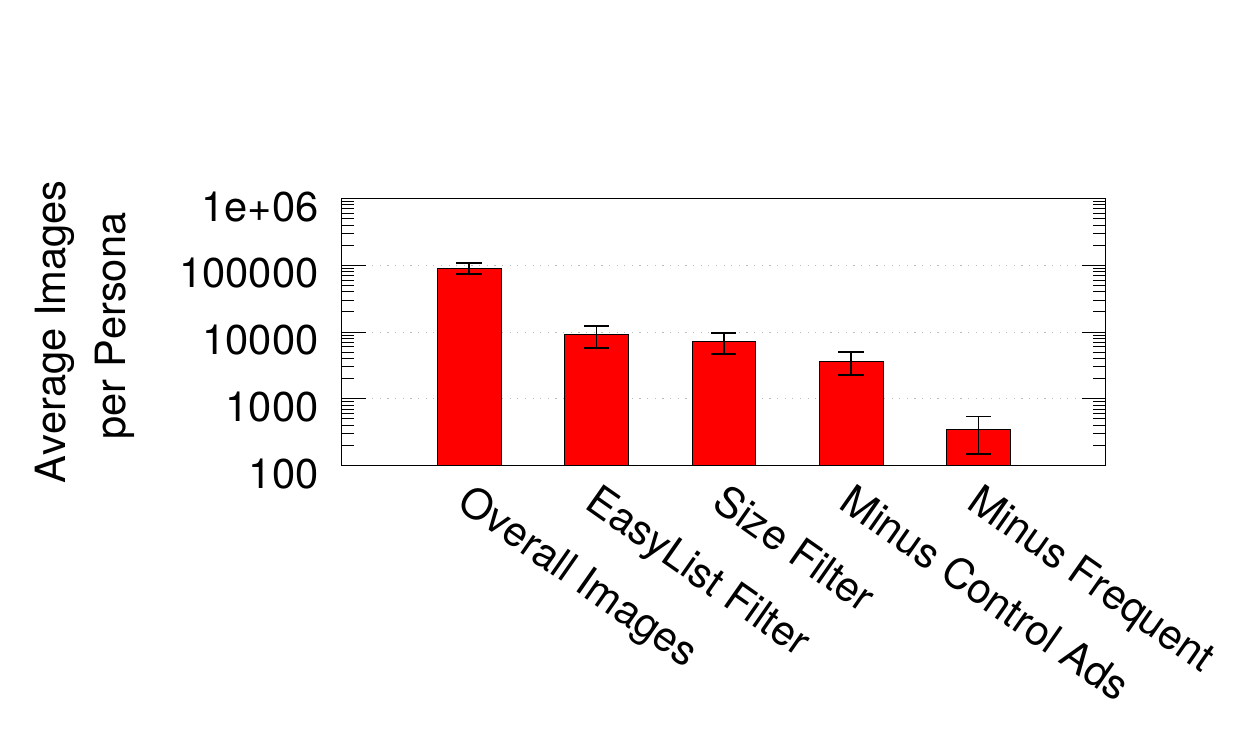}
\end{center}
\vspace{-1em}
\caption{Average number of images per persona, with standard deviation error bars.}
\label{fig:ads_per_profile}
\end{figure}

\subsection{Basic Filtering}
\label{sec:filtering}

Prior work has used a number of techniques to identify ad images from crawled data. First, we leverage the
\textit{EasyList} filter\footnote{https://easylist-downloads.adblockplus.org/easylist.txt} provided by
\textit{AdBlockPlus} to detect images that are likely to be ads~\cite{barford-adscape-2014,li-pam15}.
In our case, we look at the inclusion chain for each image, and filter out those in which none of the
URLs in the chain are a hit against EasyList. This reduces the set to 93,726 unique images.

Next, we filter out all images with dimensions $<50\times50$ pixels. These images are too small
to be ads; most are $1\times1$ tracking pixels.

Our final filter relies on a unique property of retargeted ads: they should only appear to personas that
visit a specific e-commerce site. In other words, any ad that was shown to our control account (which
visits no e-commerce sites) is either untargeted or contextually targeted, and can be discarded.
Furthermore, any ad shown to $>$1 persona may be behaviorally targeted, but it cannot be a retarget,
and is therefor filtered out\footnote{Several of our personas have retailers in common,
which we account for when filtering ads.}.

Figure~\ref{fig:ads_per_profile} shows the average number of images remaining per persona after applying each
filter. After applying all four filters, we are left with 31,850 ad images.

\subsection{Identifying Targeted \& Retargeted Ads}

At this point, we do not know which of the ad images are retargets. 
Prior work has identified retargets by looking for specific URL parameters associated with them, however this technique
is only able to identify a subset of retargets served by DoubleClick~\cite{liu-2013-AIT}. Since our goal is
to be mechanism and ad exchange agnostic, we must use a more generalizable method to identify retargets.

\para{Crowdsourcing.} Given the large number of ads in our corpus, we decided to crowdsource labels from
workers on Amazon Mechanical
Turk (AMT). We constructed Human Intelligence Tasks (HITs) that ask workers to label 30 ads, 27 of which
are unlabeled, and 3 of which are known to be retargeted ads and serve as controls (we manually identified
1,016 retargets from our corpus of 31,850 to serve as these controls).

\begin{figure}[t]
\begin{center}
\includegraphics[width=1.0\columnwidth]
{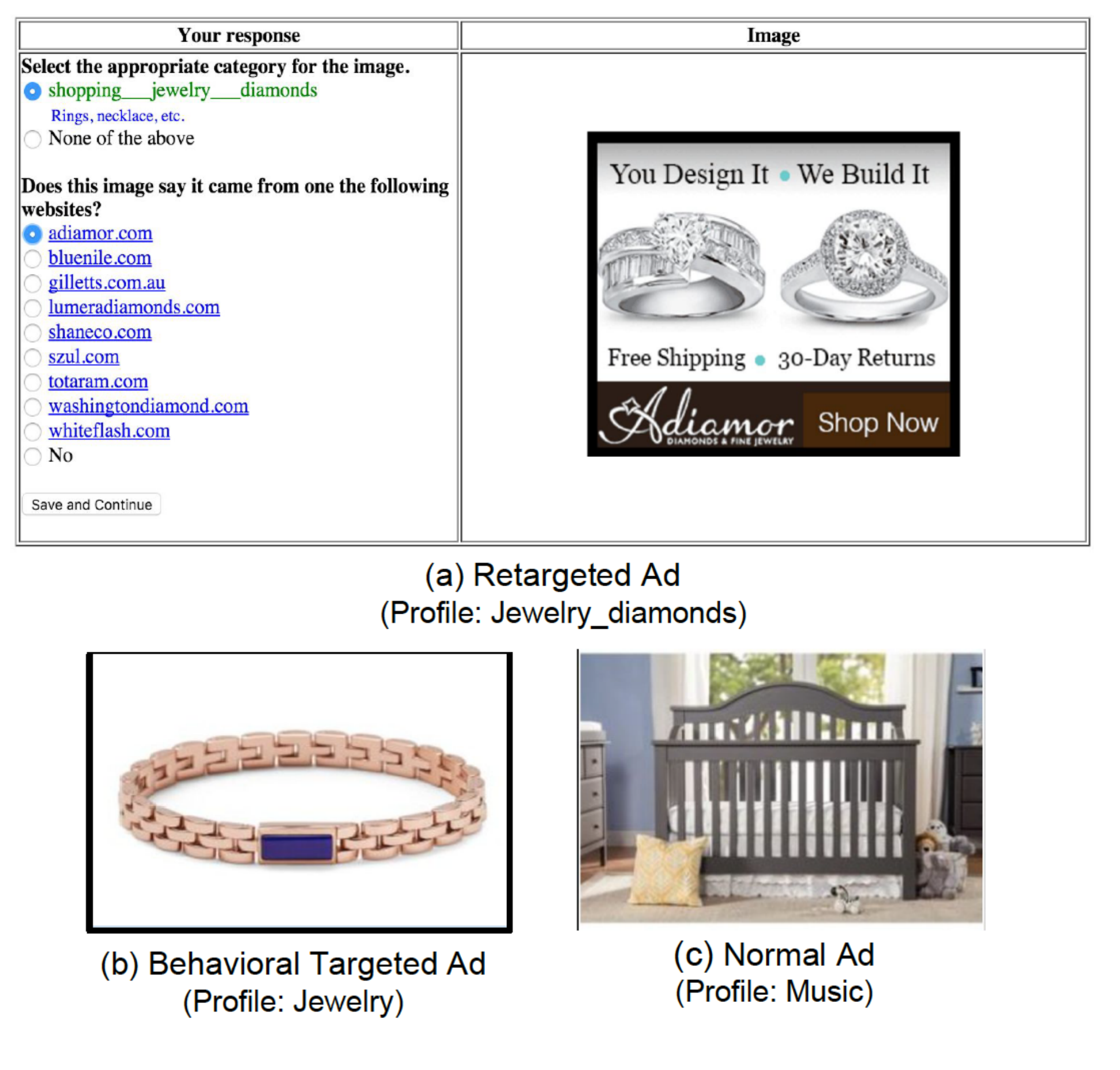}
\end{center}
\vspace{-1em}
\caption{Screenshot of our AMT HIT, and examples of different types of ads.}
\label{fig:amt_tasks}
\end{figure}

Figure~\ref{fig:amt_tasks}(a) shows a screenshot of our HIT. On the right is an ad image, and on the left
we ask the worker two questions:
\begin{packed_enumerate}
\item Does the image belong to one of the following categories (with ``None of the above'' being one option)?
\item Does the image say it came from one of the following websites (with ``No'' being one option)?
\end{packed_enumerate}
The purpose of question (1) is to isolate behavioral and retargeted ads from contextual and untargeted ads
(\eg Figure~\ref{fig:amt_tasks}(c), which was served to our {\em Music} persona).
The list for question (1) is populated with the shopping categories associated with the
persona that crawled the ad. For example, as shown in Figure~\ref{fig:amt_tasks}(a), the category
list includes ``shopping\_jewelry\_diamonds'' for ads shown to our {\em Diamond Jewelry} persona.
In most cases, this list contains exactly one entry, although there are rare cases where up to
3 categories are in the list. 

If the worker does not select ``None'' for question (1), then they are shown question (2).
Question (2) is designed to separate retargets from behavioral targeted ads. The list of websites
for question (2) is populated with the e-commerce sites visited by the persona that crawled the ad.
For example, in Figure~\ref{fig:amt_tasks}(a), the ad clearly says ``Adiamor'', and one of the sites
visited by the persona is \url{adiamor.com}; thus, this image is likely to be a retarget.

\para{Quality Control.} We apply four widely used techniques to maintain and validate the quality of
our crowdsourced image labels~\cite{wang-ndss13,hannak-2013-filterbubbles,soeller-www16}.
{\em First}, we restrict our HITs to workers that have completed $\ge$50 HITs and
have an approval rating $\ge$95\%. {\em Second}, we restrict our HITs to workers living in the US, since our
ads were collected from US websites. {\em Third}, we reject a HIT if the worker mislabels $\ge$2 of the control
images (\ie known retargeted ads); this prevents workers from being able to simply answer ``None'' to all
questions. We resubmitted rejected HITs for completion
by another worker. Overall, the workers correctly labeled 87\% of the control images. {\em Fourth} and finally,
we obtain two labels on each unlabeled image by different workers. For 92.4\% of images both labels match,
so we accept them. We manually labeled the divergent images ourselves to break the tie.

\para{Finding More Retargets.} The workers from AMT successfully identified 1,359 retargeted ads.
However, it is possible that they failed to identify some retargets, \ie there are false negatives. This
may occur in cases like Figure~\ref{fig:amt_tasks}(b): it is not clear if this ad was served as a behavioral
target based on the persona's interest in jewelry, or as a retarget for a specific jeweler.

To mitigate this issue, we manually examined all 7,563 images that were labeled as behavioral
ads by the workers. In addition to the images themselves, we also looked at the inclusion chains
for each image. In many cases, the URLs reveal that specific e-commerce sites
visited by our personas hosted the images, indicating that the ads are retargets. For example, 
Figure~\ref{fig:amt_tasks}(b) is actually part of a retargeted ad from \url{fossil.com}. 
Our manual analysis uncovered an additional 3,743 retargeted ads.

\diff{
These results suggest that the number of false negatives from our crowdsourcing task could be
dramatically reduced by showing the URLs associated with each ad image to the workers. However, note
that adding this information to the HIT will change the dynamics of the task: false negatives may go
down but the effort (and therefore the cost) of each HIT will go up. This stems from the additional
time it will take each worker to review the ad URLs for relevent keywords.

In \S~\ref{sec:analysiscookiematching}, we compare the datasets labeled by the workers and by the authors.
Interestingly, although our dataset contains a greater {\em magnitude} of retargeted ads versus the worker's
dataset, it does not improve {\em diversity}, \ie the smaller dataset identifies 96\% of the top 25 most
frequent ad networks in the larger dataset. These networks are responsible for the vast majority of
retargeted ads and inclusion chains in our dataset.
}

\para{Final Results.} Overall, we submitted 1,142 HITs to AMT. We paid \$0.18 per HIT, for a total of 
\$415. We did not collect any personal information from workers. In total, we and workers
from AMT labeled 31,850 images, of which 7,563 are behavioral targeted ads and 5,102 are retargeted ads.
These retargets advertise 281 distinct e-commerce websites (38\% of all e-commerce sites).

\subsection{Limitations}
\label{sec:limitations_labeling}

With any labeling task of this size and complexity, it is possible that there are
false positives and negatives. Unfortunately, we cannot bound these quantities, since we do not have
ground-truth information about known retargeted ad campaigns, nor is there a reliable mechanism to
automatically detect retargets (\eg based on special URL parameters, {\em etc.}).

In practice, the effect of false positives is that we will erroneously classify pairs of ad exchanges as
sharing information. \diff{We take measures to mitigate false positives by running a control crawl and removing images
which appear in multiple personas (see \S~\ref{sec:filtering}), but false positives 
can still occur.}  However, as we show in \S~\ref{sec:analysis}, the results of our classifier are extremely
consistent, suggesting that there are few false positives in our dataset.

False negatives have the opposite effect: we may miss pairs of ad exchanges that are sharing
information. Fortunately, the practical impact of false negatives is low, since we only need to correctly
identify a single retargeted ad to infer that a given pair of ad exchanges are sharing information.

\section{Analysis}
\label{sec:analysis}

In this section, we use the 5,102 retargeted ads uncovered in \S~\ref{sec:labeling}, coupled with their associated
inclusion chains (see \S~\ref{subsec:chrome}), to analyze the information flows between ad exchanges.
Specifically, we seek to answer two fundamental questions: {\em who} is sharing user data, and {\em how}
does the sharing take place (\eg client-side via cookie matching, or server-side)?

We begin by {\em categorizing} all of the retargeted ads and their associated inclusion chains into one of four
classes, which correspond to different mechanisms for sharing user data. Next, we examine specific pairs of
ad exchanges that share data, and compare our detection approach to those used in prior work to identify
cookie matching~\cite{liu-2013-AIT,acar-WNF-2014,olejnik-ndss14,falahrastegar-pam16}. We find that
prior work may be missing 31\% of collaborating
exchanges. Finally, we construct a graph that captures ad exchanges and the relationships between them,
and use it to reveal nuanced characteristics about the roles that different exchanges play in the ad ecosystem.

\begin{figure}[t]
	\centering
	\includegraphics[width=1.0\columnwidth]{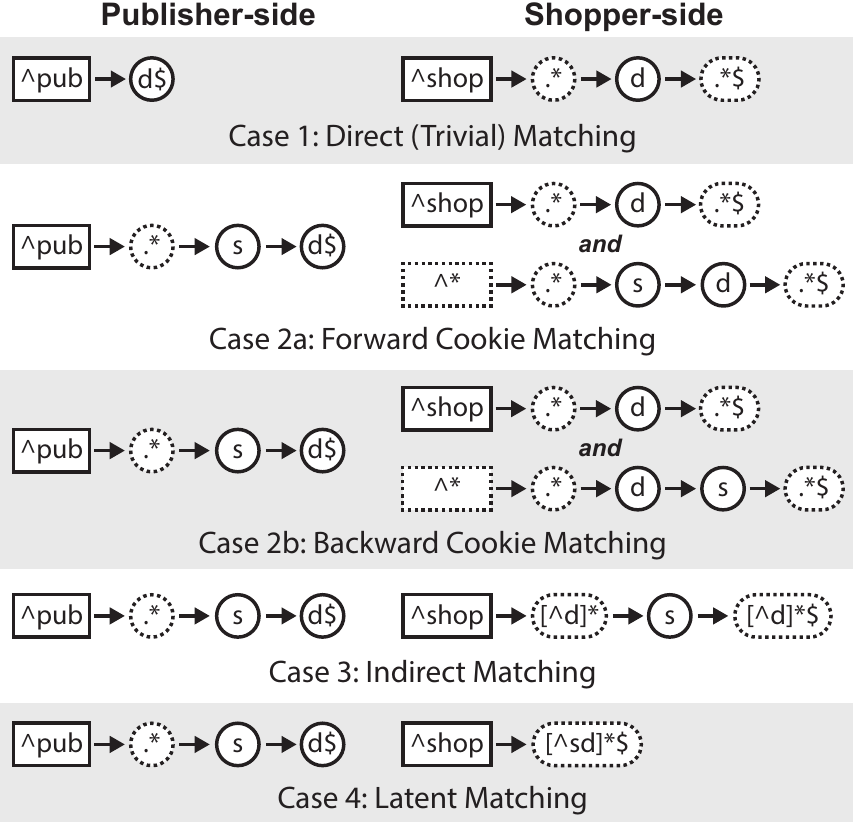}
	\caption{Regex-like rules we use to identify different types of ad exchange interactions. $shop$ and $pub$ refer to chains that begin at an e-commerce site or publisher, respectively. $d$ is the DSP that serves a retarget; $s$ is the predecessor to $d$ in the publisher-side chain, and is most likely an SSP holding an auction. Dot star ($.*$) matches any domains zero or more times.}
	\label{fig:regexchain}
\end{figure}

\subsection{Information Flow Categorization}

We begin our analysis by answering two basic questions: {\em for a given retargeted ad, was user information shared
between ad exchanges, and if so, how?} To answer these questions, we categorize the 35,448 {\em publisher-side}
inclusion chains corresponding to the 5,102 retargeted ads in our data. Note that 1) we observe some retargeted
ads multiple times, resulting in multiple chains, and 2) the chains for a given unique ad may not be identical.

We place publisher-side chains into one of four categories, each of which corresponds to a specific information
sharing mechanism (or lack thereof). To determine the category of a given chain, we match it against
carefully designed, regular expression-like rules. Figure~\ref{fig:regexchain} shows the pattern matching rules
that we use to identify chains in each category. These rules are mutually exclusive, \ie a chain will match one
or none of them.

\para{Terminology.} Before we explain each classification in detail, we first introduce shared terminology that
will be used throughout this section. Each retargeted ad was served to our persona via a {\em publisher-side} chain.
$pub$ is the domain
of the publisher at the root of the chain, while $d$ is the domain at the end of the chain that served the ad.
Typically, $d$ is a DSP. If the retarget was served via an auction, then an SSP $s$ must immediately precede $d$ 
in the publisher-side chain.

Each retarget advertises a particular e-commerce site. $shop$ is the domain of the e-commerce site corresponding
to a particular retargeted ad. To categorize a given publisher-side chain, we must also consider the corresponding
{\em shopper-side} chains rooted at $shop$.

\subsubsection{Categorization Rules}
\label{sec:cat_rules}

\para{Case 1: Direct Matches.} The first chain type that we define are {\em direct matches}. Direct matches are
the simplest type of chains that can be used to serve a retargeted ad. As shown in Figure~\ref{fig:regexchain},
for us to categorize a publisher-side chain as a direct match, it must be exactly length two, with a direct
resource inclusion request from $pub$ to $d$. $d$ receives any cookies they have stored on the persona inside this
request, and thus it is trivial for $d$ to identify our persona. 

On the shopper-side, the only requirement is that $d$ observed our persona browsing $shop$. If $d$ does not observe
our persona at $shop$, then $d$ would not serve the persona a retargeted ad for $shop$.
$d$ is able to set a cookie on our persona, allowing $d$ to re-identify the persona in future.

We refer to direct matching chains as ``trivial'' because it is obvious how $d$ is able to track our persona and
serve a retargeted ad for $shop$. Furthermore, in these cases no user information needs to be shared between ad
exchanges, since there are no ad auctions being held on the publisher-side.

\para{Case 2: Cookie Matching.} The second chain type that we define are {\em cookie matches}. As the name implies,
chains in this category correspond to instance where an auction is held on the publisher-side, and we observe
direct resource inclusion requests between the SSP and DSP, implying that they are matching cookies.

As shown in Figure~\ref{fig:regexchain}, for us to categorize a publisher-side chain as cookie matching, $s$ and
$d$ must be adjacent at the end of the chain. On the shopper-side, $d$ must observe the persona at $shop$. Lastly, we
must observe a request from $s$ to $d$ or from $d$ to $s$ in
some chain before the retargeted ad is served. These requests capture the moment when the two ad exchanges match
their cookies. Note that $s \rightarrow d$ or $d \rightarrow s$ can occur in a publisher- or shopper-side chain;
in practice, it often occurs in a chain rooted at $shop$, thus fulfilling both requirements at once.

For the purposes of our analysis, we distinguish between {\em forward} ($s \rightarrow d$) and {\em backward}
($d \rightarrow s$) cookie matches. Figure~\ref{fig:cookiematch} shows an example of a forward cookie match. As
we will see, many pairs of ad exchanges engage
in both forward and backward matching to maximize their opportunities for data sharing. To our knowledge,
no prior work examines the distinction between forward and backward cookie matching.

\para{Case 3: Indirect Matching.} The third chain type we define are {\em indirect matches}. Indirect matching
occurs when an SSP sends meta-data about a user to a DSP, to help them determine if they should
bid on an impression. With respect to retargeted ads, the SSP tells the DSPs about the browsing
history of the user, thus enabling the DSPs to serve retargets for specific retailers, even if the DSP never
directly observed the user browsing the retailer (hence the name, {\em indirect}). Note that no cookie matching
is necessary in this case for DSPs to serve retargeted ads.

As shown in Figure~\ref{fig:regexchain}, the crucial difference between cookie matching chains and indirect chains
is that $d$ {\em never} observes our persona at $shop$; only $s$ observes our persona at $shop$. Thus, by inductive
reasoning, we must conclude that $s$ shares information about our persona with $d$, otherwise $d$ would never serve
the persona a retarget for $shop$.

\para{Case 4: Latent Matching.} The fourth and final chain type that we define are {\em latent matches}. As shown
in Figure~\ref{fig:regexchain}, the defining characteristic of latent chains is that neither $s$ nor $d$ observe
our persona at $shop$. This begs the question: how do $s$ and $d$ know to serve a retargeted ad for $shop$ if they
never observe our persona at $shop$? The most reasonable explanation is that some other ad exchange $x$ that is present
in the shopper-side chains shares this information with $d$ behind-the-scenes. 

We hypothesize that the simplest way for ad exchanges to implement latent matching is by having $x$ and $d$ share the
same unique identifiers for users. Although $x$ and $d$ are different domains, and are thus prevented by the SOP from
reading each others' cookies, both ad exchanges may use the same deterministic algorithm for generating user IDs
(\eg by relying on IP addresses or browser fingerprints). However, as we will show, these synchronized identifiers
are not necessarily visible from the client-side (\ie the values of cookies set by $x$ and $d$ may be obfuscated),
which prevents trivial identification of latent cookie matching.

\diff{
\para{Note:} Although we do not expect to see cases 3 and 4, they can still occur. We explain in \S~\ref{sec:catresults}
that indirect and latent matching is mostly performed by domains belonging to the same company. The remaining few instances
of these cases are probably mislabeled behaviorally targeted ads.
}

\subsubsection{Categorization Results}
\label{sec:catresults}

\begin{table}[t]
\centering
\footnotesize
\rowcolors{2}{gray!10}{white}
\begin{tabular}{r|rr|rr}
 & \multicolumn{2}{c|}{\textbf{Unclustered}} & \multicolumn{2}{c}{\textbf{Clustered}} \\
\textbf{Type} & \textbf{Chains} & \textbf{\%} & \textbf{Chains} & \textbf{\%} \\
\hline
Direct & 1770 & 5\% & 8449 & 24\% \\
Forward Cookie Match & 24575 & 69\% & 25873 & 73\% \\
Backward Cookie Match & 19388 & 55\% & 24994 & 70\% \\
Indirect Match & 2492 & 7\% & 178 & 1\% \\
Latent Match & 5362 & 15\% & 343 & 1\% \\
\hline
{\em No Match} & 775 & 2\% & 183 & 1\% \\
\end{tabular}
\caption{Results of categorizing publisher-side chains, before and after clustering domains.}
\label{table:chain_categorization}
\end{table}

\begin{table*}[t]
\centering
\footnotesize
\rowcolors{2}{gray!10}{white}
\begin{tabular}{r@{\hskip 0.70em}c@{\hskip 0.70em}l|cc|cc|cc}
 &  &  & \multicolumn{2}{c|}{\textbf{All Data}}  & \multicolumn{2}{c|}{\textbf{AMT Only}} & & \\
\textbf{Participant 1} &  & \textbf{Participant 2} & \textbf{Chains} & \textbf{Ads} & \textbf{Chains} & \textbf{Ads} & \multicolumn{2}{c}{\textbf{Heuristics}} \\
\hline
criteo & $\leftrightarrow$ & googlesyndication & 9090 & 1887  & 1629 & 370 & \multicolumn{2}{c}{$\leftrightarrow$: US} \\
criteo & $\leftrightarrow$ & doubleclick & 3610 & 1144 & 770 & 220 & $\rightarrow$: E, US & $\leftarrow$: DC, US \\
criteo & $\leftrightarrow$ & adnxs & 3263 & 1066  & 511 & 174 & \multicolumn{2}{c}{$\leftrightarrow$: E, US} \\
criteo & $\leftrightarrow$ & googleadservices & 2184 & 1030  & 448 & 214 & $\rightarrow$: E, US & $\leftarrow$: US \\
criteo & $\leftrightarrow$ & rubiconproject & 1586 & 749 & 240 & 113 & \multicolumn{2}{c}{$\leftrightarrow$: E, US}\\
criteo & $\leftrightarrow$ & servedbyopenx & 707 & 460 & 111 & 71 & \multicolumn{2}{c}{$\leftrightarrow$: US}\\
mythings & $\leftrightarrow$ & mythingsmedia & 478 & 52 & 53 & 1 & $\rightarrow$: E, US & $\leftarrow$: US \\
criteo & $\leftrightarrow$ & pubmatic & 363 & 246 &  64 & 37 & $\rightarrow$: E, US & $\leftarrow$: US\\
doubleclick & $\leftrightarrow$ & steelhousemedia & 362 & 27  & 151 & 16 & $\rightarrow$: US & $\leftarrow$: E, US \\
mathtag & $\leftrightarrow$ & mediaforge & 360 & 124 & 63 & 13 & \multicolumn{2}{c}{$\leftrightarrow$: E, US}\\
netmng & $\leftrightarrow$ & scene7 & 267 & 162 & 45 & 32 & $\rightarrow$: E & $\leftarrow$: -  \\
criteo & $\leftrightarrow$ & casalemedia & 200 & 119 & 54 & 31  & $\rightarrow$: E, US & $\leftarrow$: US\\
doubleclick & $\leftrightarrow$ & googlesyndication & 195 & 81 & 101 & 62 & \multicolumn{2}{c}{$\leftrightarrow$: US} \\
criteo & $\leftrightarrow$ & clickfuse & 126 & 99 & 14 & 13 & \multicolumn{2}{c}{$\leftrightarrow$: US}  \\
criteo & $\leftrightarrow$ & bidswitch & 112 & 78 & 25 & 15 & $\rightarrow$: E, US & $\leftarrow$: US  \\
googlesyndication & $\leftrightarrow$ & adsrvr & 107 & 29 & 102 & 24 & \multicolumn{2}{c}{$\leftrightarrow$: US} \\
rubiconproject & $\leftrightarrow$ & steelhousemedia & 86 & 30 & 43 & 19 & \multicolumn{2}{c}{$\leftrightarrow$: E}  \\
amazon-adsystem & $\leftrightarrow$ & ssl-images-amazon & 98 & 33 & 33 & 7 & \multicolumn{2}{c}{-} \\
googlesyndication & $\leftrightarrow$ & steelhousemedia & 47 & 22 & 36 & 16 & \multicolumn{2}{c}{-} \\
adtechus & $\rightarrow$ & adacado & 36 & 18 & 36 & 18 & \multicolumn{2}{c}{-}  \\
googlesyndication & $\leftrightarrow$ & 2mdn & 40 & 19 & 39 & 18 & $\rightarrow$: US & $\leftarrow$: -  \\
atwola & $\rightarrow$ & adacado & 32 & 6 & 28 & 5 & \multicolumn{2}{c}{-} \\
adroll & $\leftrightarrow$ & adnxs & 31 & 8 & 26 & 7 & \multicolumn{2}{c}{-} \\
googlesyndication & $\leftrightarrow$ & adlegend & 31 & 22 & 29 & 20 & \multicolumn{2}{c}{-} \\
adnxs & $\leftrightarrow$ & esm1 & 46 & 1 & 0 & 0 & $\rightarrow$: US & $\leftarrow$: - \\
\end{tabular}
\caption{Top 25 cookie matching partners in our dataset. The arrow signifies whether we observe 
forward matches ($\rightarrow$), backward matches ($\leftarrow$), or both ($\leftrightarrow$). 
The heuristics for detecting cookie matching are: {\em DC} (match using DoubleClick URL parameters), {\em E}
(string match for exact cookie values), {\em US} (URLs that include parameters like ``usersync''), and
- (no identifiable mechanisms). Note that the HTTP request formats used for forward and backward matches
between a given pair of exchanges may vary.}
\label{table:cookie-partners}
\end{table*}

We applied the rules in Figure~\ref{fig:regexchain} to all 35,448 publisher-side chains in our dataset twice.
First, we categorized the raw, unmodified chains; then we {\em clustered} domains that belong to the same
companies, and categorized the chains again. For example, Google owns \url{youtube.com},
\url{doubleclick.com}, and \url{2mdn.net}; in the clustered experiments,
we replace all instances of these domains with \url{google.com}. Appendix~\ref{app:clusters} lists all clustered domains.

Table~\ref{table:chain_categorization} presents the results of our categorization. The first thing we observe
is that cookie matching is the most frequent classification by a large margin.
This conforms to our expectations, given that RTB is widespread in today's ad ecosystem, and major exchanges like
DoubleClick support it~\cite{DC-RTB}. Note that, for a given ($s$, $d$) pair in a publisher-side
chain, we may observe $s \rightarrow d$ and $d \rightarrow s$ requests in our data, \ie the pair engages in forward
and backward cookie matching. This explains why the percentages in Table~\ref{table:chain_categorization} do not
add up to 100\%.

The next interesting feature that we observe in Table~\ref{table:chain_categorization} is that indirect and latent
matches are relatively rare (7\% and 15\%, respectively). Again, this is expected, since these types
of matching are more exotic and require a greater degree of collaboration between ad exchanges to implement.
Furthermore, the percentage of indirect and latent matches drops to 1\% when we cluster domains. To understand
why this occurs, consider the following real-world example chains:

\vspace{0.5em}
\parab{Publisher-side:} $pub \rightarrow rubicon \rightarrow googlesyndication$

\parab{Shopper-side:} $shop \rightarrow doubleclick$
\vspace{0.5em}

\noindent According to the rules in Figure~\ref{fig:regexchain}, this appears to be a latent match, since Rubicon
and Google Syndication do not observe our persona on the shopper-side. However, after clustering the Google
domains, this will be classified as cookie matching (assuming that there exists at least one other request from
Rubicon to Google).

The above example is extremely common in our dataset: 731 indirect chains become cookie matching
chains after we cluster the Google domains {\em alone}. Importantly, this finding provides strong evidence that Google
does in fact use latent matching to share user tracking data between its various domains. Although this is allowed
in Google's terms of service as of 2014~\cite{google-tos}, our results provide direct evidence of this data sharing
with respect to serving targeted ads. In the vast majority of these cases, Google Syndication is the DSP, suggesting
that on the server-side, it ingests tracking data and user identifiers from all other Google services (\eg DoubleClick
and Google Tag Manager).

Of the remaining 1\% of chains that are still classified as indirect or latent after clustering, the majority appear
to be false positives. In most of these cases, we observe $s$ and $d$ doing cookie matching in other instances, and it
seems unlikely that $s$ and $d$ would also utilize indirect and latent mechanisms. These ads are probably
mislabeled behaviorally targeted ads.

The final takeaway from Table~\ref{table:chain_categorization} is that the number of uncategorized
chains that do not match any of our rules is extremely low (1-2\%). These publisher-side chains are likely to
be false positives, \ie ads that are not actually retargeted. These results suggest that our image labeling
approach is very robust, since the vast majority of chains are properly classified as direct or cookie matches.

\begin{table*}[t]
\centering
\footnotesize
\rowcolors{2}{gray!10}{white}
\begin{tabular}{rrr|ccc|ccc|cc}
& & & \multicolumn{3}{c|}{\textbf{Degree}} & \multicolumn{3}{c|}{\textbf{Position $p$ in Chains (\%)}} & \textbf{\# of Shopper} &  \\
 & & \textbf{Domain} & \textbf{In} & \textbf{Out} & \textbf{In/Out Ratio} & \textbf{$p_2$} & \textbf{$p_{n-1}$} & \textbf{$p_n$} & \textbf{Websites} & \textbf{\# of Ads} \\
\hline
& & criteo & 35 & 6 & 5.83 & 9.28 & 0.00 & 68.8 & 248 & 3,335 \\
& & mediaplex & 8 & 2 & 4.00 & 0.00 & 85.7 & 0.07 & 20  & 14 \\
& & tellapart & 6 & 1 & 6.00 & 25.0 & 100.0 & 0.18 & 33  & 9 \\
& & mathtag & 12 & 6 & 2.00 & 0.00 & 90.9 & 0.06 & 314 & 2 \\
& & mythingsmedia & 1 & 0 & - & 0.00 & 0.00 & 1.41 & 1 & 59 \\
& & steelhousemedia & 8 & 0 & - & 0.00 & 0.00 & 16.8 & 40  & 89 \\
\parbox[t]{2mm}{\multirow{-7}{*}{\rotatebox[origin=c]{90}{\textbf{DSPs}}}} & & mediaforge & 5 & 0 & - & 0.00 & 0.00 & 1.28 & 29 & 143 \\
\hline
& & pubmatic & 5 & 9 & 0.56 & 3.17 & 74.2 & 0.01 & 362 & 4 \\
& & rubiconproject & 19 & 22 & 0.86 & 23.5 & 62.8 & 0.01 & 394 & 3 \\
& & adnxs & 18 & 20 & 0.90 & 94.2 & 91.9 & 0.16 & 476 & 12 \\
& & casalemedia & 9 & 10 & 0.90 & 1.30 & 90.0 & 0.00 & 298 & 0 \\
\hhline{~----------}
\hhline{~~~~~~~~~~~}
& & atwola & 4 & 19 & 0.21 & 84.6 & 18.2 & 0.01 & 62 & 2 \\
& & advertising & 4 & 4 & 1.00 & 0.00 & 75.0 & 0.10 & 337 & 17 \\
& \parbox[t]{2mm}{\multirow{-3}{*}{\rotatebox[origin=c]{90}{AOL}}} & adtechus & 17 & 16 & 1.06 & 1.58 & 27.3 & 0.09 & 328 & 15 \\
\hhline{~----------}
\hhline{~~~~~~~~~~~}
& & servedbyopenx & 6 & 11 & 0.55 & 7.2 & 83.8 & 0.00 & 2 & 0 \\
& & openx & 10 & 9 & 1.11 & 0.95 & 9.83 & 0.00 & 390 & 0 \\
\parbox[t]{2mm}{\multirow{-10}{*}{\rotatebox[origin=c]{90}{\textbf{SSPs}}}} & \parbox[t]{2mm}{\multirow{-3}{*}{\rotatebox[origin=c]{90}{OpenX}}} & openxenterprise & 4 & 4 & 1.00 & 40.0 & 20.0 & 0.00 & 1 & 0 \\
\hline
& & googletagservices & 44 & 2 & 22.00 & 93.7 & 0.00 & 0.00 & 65 & 0 \\
& & googleadservices & 4 & 17 & 0.24 & 2.94 & 33.5 & 0.00 & 485 & 0 \\
& & 2mdn & 3 & 1 & 3.00 & 0.00 & 0.00 & 1.35 & 62 & 125 \\
& & googlesyndication & 90 & 35 & 2.57 & 70.1 & 62.7 & 19.8 & 84 & 638 \\
\parbox[t]{2mm}{\multirow{-5}{*}{\rotatebox[origin=c]{90}{\textbf{Google}}}} & & doubleclick & 38 & 36 & 1.06 & 38.8 & 63.1 & 0.22 & 675 & 19 \\
\end{tabular}
\caption{Overall statistics about the connectivity, position, and frequency of ad domains in our dataset.}
\label{table:popular_retargeters}
\end{table*}

\subsection{Cookie Matching}
\label{sec:analysiscookiematching}

The results from the previous section confirm that cookie matching is ubiquitous on
today's Web, and that this information sharing fuels highly targeted advertisements.
Furthermore, our classification results demonstrate that we can detect cookie matching
without relying on semantic information about cookie matching mechanisms.

In this section, we take a closer look at the pairs of ad exchanges that we observe matching
cookies. We seek to answer two questions: {\em first}, which pairs match most frequently,
and what is the directionality of these relationships? {\em Second}, what fraction of cookie
matching relationships will be missed by the heuristic detection approaches used by prior
work~\cite{liu-2013-AIT,acar-WNF-2014,olejnik-ndss14,falahrastegar-pam16}?

\para{Who Is Cookie Matching?} Table~\ref{table:cookie-partners} shows the top 25 most
frequent pairs of domains that we observe matching cookies. The arrows indicate the direction
of matching (forward, backward, or both). ``Ads'' is the number of unique retargets served
by the pair, while ``Chains'' is the total number of associated publisher-side
chains. \diff{We present both quantities as observed in our complete dataset (containing 5,102
retargets), as well as the subset that was identified solely by the AMT workers
(containing 1,359 retargets).}

We observe that cookie matching frequency is heavily skewed towards several heavy-hitters.
In aggregate, Google's domains are most common, which makes sense given that Google
is the largest ad exchange on the Web today. The second most common is Criteo; this
result also makes sense, given that Criteo specializes in retargeted advertising~\cite{criteo-econsultancy}. 
\diff{These observations remain broadly true across the AMT and complete datasets: although the
relative proportion of ads and chains from less-frequent exchange pairs differs somewhat
between the two datasets, the heavy-hitters do not change. Furthermore, we also see
that the vast majority of exchange pairs are identified in both datasets.}

Interestingly, we observe a great deal of heterogeneity with respect to the directionality
of cookie matching. Some boutique exchanges, like Adacado, only ingest cookies from other
exchanges. Others, like Criteo, are omnivorous, sending or receiving data from any and all
willing partners. These results suggest that some participants are more wary about releasing
their user identifiers to other exchanges.

\para{Comparison to Prior Work.} We observe many of the same participants matching cookies as
prior work, including DoubleClick, Rubicon, AppNexus, OpenX, MediaMath, and
myThings~\cite{acar-WNF-2014,olejnik-ndss14,falahrastegar-pam16}. Prior
work identifies some additional ad exchanges that we do not (\eg Turn); this is due to our
exclusive focus on participants involved in retargeted advertising.

However, we also observe participants (\eg Adacado and AdRoll) that prior work does not.
This may be because prior work identifies cookie matching using heuristics to pick out specific
features in HTTP requests~\cite{liu-2013-AIT,acar-WNF-2014,olejnik-ndss14,falahrastegar-pam16}.
In contrast, our categorization approach is content and mechanism agnostic.

To investigate the efficacy of heuristic detection methods, we applied three of them to our
dataset. Specifically, for each pair ($s$, $d$) of exchanges that we categorize as cookie matching, we
apply the following tests to the HTTP headers of requests between $s$ and $d$ or vice-versa:
\begin{packed_enumerate}
\item We look for specific keys that are known to be used by DoubleClick and other Google
domains for cookie matching (\eg ``google\_nid''~\cite{olejnik-ndss14}).
\item We look for cases where unique cookie values set by one participant are included in requests
sent to the other participant\footnote{To reduce false positives, we only consider cookie values that have
length $>$10 and $<$100.}.
\item We look for keys with revealing names like ``usersync'' that frequently appear in requests
between participants in our data.
\end{packed_enumerate}

As shown in the ``Heuristics'' column in Table~\ref{table:cookie-partners},
in the majority of cases, heuristics are able to identify cookie matching between the participants.
Interestingly, we observe that the mechanisms used by some pairs (\eg Criteo and
DoubleClick) change depending on the
directionality of the cookie match, revealing that the two sides have different cookie matching APIs.

However, for 31\% of our cookie matching partners, the heuristics are unable to detect signs
of cookie matching. We hypothesize that this is due to obfuscation techniques employed by specific
ad exchanges. In total, there are 4.1\% cookie matching chains that would be completely missed by
heuristic tests. This finding highlights the limitations of prior work, and
bolsters the case for our mechanism-agnostic classification methodology.

\subsection{The Retargeting Ecosystem}

In this last section, we take a step back and examine the broader ecosystem for retargeted ads that is revealed by
our dataset. To facilitate this analysis, we construct a graph by taking the union of all of our publisher-side chains.
In this graph, each node is a domain (either a publisher or an ad exchange), and edges correspond to resource inclusion
relationships between the domains. Our graph formulation differs from prior work in that edges denote actual information
flows, as opposed to simple co-occurrences of trackers on a given domain~\cite{gomer-wiiat13}.

Table~\ref{table:popular_retargeters} presents statistics on the top ad-related domains in our dataset. The
``Degree'' column shows the in- and out-degree of nodes, while ``Position'' looks at the relative location of
nodes within chains. $p_2$ is the second position in the chain, corresponding to the first ad network after the publisher;
$p_n$ is the DSP that serves the retarget in a chain of length $n$; $p_{n-1}$ is the second to last position,
corresponding to the final SSP before the DSP. Note that a domain may appear in a chain multiple times, so the sum of
the $p_i$ percentages may be $>$100\%. The last two columns count the number of unique e-commerce sites that embed
resources from a given domain, and the unique number of ads served by the domain.

Based on the data in Table~\ref{table:popular_retargeters}, we can roughly cluster the ad domains into two groups,
corresponding to SSPs and DSPs. DSPs have low or zero out-degree since they often appear at position $p_n$, \ie
they serve an ad and terminate the chain. Criteo is the largest source of retargeted ads in our dataset
by an order of magnitude. \diff{This is not surprising, given that Criteo was identified as the largest retargeter 
in the US and UK in 2014~\cite{criteo-econsultancy}.}

In contrast, SSPs tend to have in/out degree ratios closer to 1, since they facilitate the exchange of ads between
multiple publishers, DSPs, and even other SSPs. Some SSPs, like Atwola, work more closely with publishers and thus
appear more frequently at $p_2$, while others, like Mathtag, cater to other SSPs and thus appear almost exclusively
at $p_{n-1}$. Most of the SSPs we observe also function as DSPs (\ie they serve some retargeted ads), but there are
``pure'' SSPs like Casale Media and OpenX that do not serve ads. Lastly,
Table~\ref{table:popular_retargeters} reveals that SSPs tend to do more user tracking than DSPs, by getting
embedded in more e-commerce sites (with Criteo being the notable exception).

Google is an interesting case study because its different domains have clearly delineated purposes.
\texttt{googletagservices} is Google's in-house SSP, which funnels impressions directly from publishers to
Google's DSPs: \texttt{2mdn}, \texttt{googlesyndication}, and \texttt{doubleclick}. In contrast,
\texttt{googleadservices} is also an SSP, but it holds auctions with third-party participants (\eg Criteo).
\texttt{googlesyndication} and \texttt{doubleclick} function as both SSPs and DSPs, sometimes holding auctions,
and sometimes winning auctions held by others to serve ads. Google Syndication is the second most frequent
source of retargeted ads in our dataset behind Criteo.

\section{Concluding Discussion}
\label{sec:conclusion}

In this study, we develop a novel, principled methodology for detecting flows of tracking information
between ad exchanges. The key insight behind our approach is that we re-purpose retargeted ads as
a detection mechanism, since their presence reveals information flows between ad exchanges.
Our methodology is content-agnostic, and thus we are able to identify flows even
if they occur on the server-side. This is a significant improvement over prior work, which relies
on heuristics to detect cookie matching~\cite{acar-WNF-2014,olejnik-ndss14,falahrastegar-pam16}.
As we show in \S~\ref{sec:analysis}, these heuristics fail to detect 31\% of matching pairs
today, and they are likely to fail more in the future as ad networks adopt content
obfuscation techniques.

\para{Implications for Users.} Ultimately, our goal is not just to measure information flows
between ad exchanges, but to facilitate the development of systems that balance user privacy
against the revenue needs of publishers.

Currently, users are faced with unsatisfactory choices when deciding if and how to block ads and
tracking. Whitelisting approaches like NoScript are effective at protecting privacy, but are too
complicated for most users, and deprive publishers of revenue. Blocking third-party cookies is
ineffective against first-party trackers (\eg Facebook). AdBlockPlus' controversial
``Acceptable Ads'' program is poorly governed and leaves users vulnerable to unscrupulous ad
networks~\cite{walls-imc15}. DNT is DOA~\cite{balebako-w2sp12}. Although researchers have proposed
privacy preserving ad exchanges, these systems have yet to see widespread
adoption~\cite{fredrikson-oakland11,guha-nsdi11,backes-oakland12}.

\diff{
We believe that data about information flows between ad exchanges potentially opens up a new
middle ground in ad blocking. One possibility is to develop an automated system that uses the
methodology developed in this paper to continuously crawl ads, identify cookie matching flows,
and construct rules that match these flows. Users could then install a browser extension that
blocks flows matching these rules. The advantage of this extension is that it
would offer improved privacy protection relative to existing systems (\eg Ghostery and
Disconnect), while also allowing advertising (as opposed to traditional ad blockers).
However, the open challenge with this system design would be making it cost effective, since it
would still rely crowdsourced labor.

Another possibility is using our data as ground-truth for a
sophisticated blocker that relies on client-side Information Flow Control (IFC). There
exist many promising, lightweight approaches to implementing JavaScript IFC in the
browser~\cite{hedin-sac14,bichhawat-2014,stefan-osdi14,heule-post15}.
However, IFC alone is not enough to block cookie matching flows: as we have shown, ad networks
obfuscate data, making it impossible to separate benign from ``leaky'' flows in general. Instead,
we can use information gathered using our methodology as ground-truth to mark data in
specific incoming flows, and rely on IFC to enforce restrictions that prevent outgoing flows
from containing the marked data.
}

\section*{Acknowledgements}

We thank our shepherd, Nektarios Leontiadis, and the anonymous reviewers for their helpful comments.  This research was supported in part by NSF grants CNS-1319019 and CHS-1408345. Any opinions, findings, and conclusions or recommendations expressed in this material are those of the authors and do not necessarily reflect the views of the NSF.

\balance

{
\footnotesize
\bibliographystyle{acm}
\bibliography{adretargeting}
}

\appendix
\section{Appendix}

\subsection{Clustered Domains}
\label{app:clusters}

We clustered the following domains together when classifying publisher-side chains
in \S~\ref{sec:catresults}.

\parab{Google:} google-analytics, googleapis, google, doubleclick, gstatic, googlesyndication,
googleusercontent, googleadservices, googletagmanager, googletagservices, googlecommerce,
youtube, ytimg, youtube-mp3, googlevideo, 2mdn

\parab{OpenX:} openxenterprise, openx, servedbyopenx

\parab{Affinity:} affinitymatrix, affinity

\parab{Ebay:} ebay, ebaystatic

\parab{Yahoo:} yahoo, yimg

\parab{Mythings:} mythingsmedia, mythings

\parab{Amazon:} cloudfront, amazonaws, amazon-adsystem, images-amazon

\parab{Tellapart:} tellapart, tellaparts

\end{document}